\documentclass{elsart}
\journal{Physics Letter B}
\usepackage{graphicx}
\usepackage{color}

\makeatletter
\DeclareRobustCommand*\cal{\@fontswitch\relax\mathcal}

\begin{document}

\begin{frontmatter}

\begin{flushleft}
\hspace*{9cm}BELLE Preprint 2009-28 \\
\hspace*{9cm}KEK Preprint 2009-37 \\
\hspace*{9cm}NTLP Preprint 2009-05 \\
\end{flushleft}

\title{
Search for Lepton Flavor Violating $\tau$ Decays into Three Leptons
with 719 Million Produced $\tau^+\tau^-$ Pairs
}

\collab{Belle Collaboration}
  \author[Nagoya]{K.~Hayasaka}, 
  \author[Nagoya]{K.~Inami}, 
  \author[Nagoya]{Y.~Miyazaki}, 
  \author[BINP,Novosibirsk]{K.~Arinstein}, 
  \author[BINP,Novosibirsk]{V.~Aulchenko}, 
  \author[Lausanne,ITEP]{T.~Aushev}, 
  \author[Sydney]{A.~M.~Bakich}, 
  \author[Lausanne]{A.~Bay}, 
  \author[Protvino]{K.~Belous}, 
  \author[Panjab]{V.~Bhardwaj}, 
% ...

\address[BINP]{Budker Institute of Nuclear Physics, Novosibirsk, Russian Federation}
% ...
\address[Protvino]{Institute for High Energy Physics, Protvino, Russian Federation}
% ...
\address[ITEP]{Institute for Theoretical and Experimental Physics, Moscow, Russian Federation}
% ...
\address[Lausanne]{\'Ecole Polytechnique F\'ed\'erale de Lausanne, EPFL, Lausanne, Switzerland}
% ...
\address[Nagoya]{Nagoya University, Nagoya, Japan}
% ...
\address[Novosibirsk]{Novosibirsk State University, Novosibirsk, Russian Federation}
% ...
\address[Panjab]{Panjab University, Chandigarh, India}
% ...
\address[Sydney]{School of Physics, University of Sydney, NSW 2006, Australia}
% ...

\begin{abstract}
We present a search for lepton-flavor-violating $\tau$ decays 
into three leptons (electrons or muons) using 782 fb$^{-1}$ of data collected 
with the Belle detector at the KEKB asymmetric-energy $e^+e^-$ collider. 
No evidence for these decays is observed and 
we set 90\% confidence level upper limits 
on the branching fractions between 
$1.5\times 10^{-8}$ and $2.7\times 10^{-8}$.
\end{abstract}

\end{frontmatter}

\section{Introduction}

Lepton flavor violation (LFV) appears
in various extensions of the Standard Model (SM).
In particular, lepton-flavor-violating
$\tau^-\to\ell^-\ell^+\ell^-$ (where $\ell = e$ or $\mu$ )
decays are discussed in 
various supersymmetric models~\cite{cite:susy1,cite:susy2,cite:susy3,cite:susy4,cite:susy5,cite:susy6,cite:susy7,cite:susy8},
models with
little Higgs~\cite{cite:littlehiggs1,cite:littlehiggs2},
left-right symmetric models~\cite{cite:leftright}
as well as models with
heavy singlet Dirac neutrinos~\cite{cite:amon}
and
very light pseudoscalar bosons~\cite{cite:pseudo}.
Some of these models with certain combinations of parameters 
predict that the branching fractions 
for $\tau^-\to\ell^-\ell^+\ell^-$ decays
can be as large as $10^{-7}$, which is in the range
already accessible in high-statistics  $B$ factory experiments.

Searches for lepton flavor violation in
$\tau^- \to \ell^- \ell^+ \ell^-$ (where $\ell = e$ or $\mu$) decays
have been performed since 1982~\cite{PDG}, starting
from the pioneering experiment MARKII~\cite{MARKII}.
In the previous high-statistics analyses,
Belle (BaBar) reached 90\% confidence level upper limits 
on the branching fractions of the 
order of $10^{-8}$~\cite{cite:3l_belle,cite:3l_babar}, based on 
samples with about 535 (376) fb${}^{-1}$ of data.
Here, we update our previous results 
with a larger data set (782 fb$^{-1}$), collected with the Belle detector 
at the KEKB asymmetric-energy $e^+e^-$ collider~\cite{kekb},
taken at the $\Upsilon(4S)$ resonance and 60 MeV below it.
We apply the same selection criteria as in the previous
analysis, but optimized for the new data sample.

The Belle detector is a large-solid-angle magnetic spectrometer that
consists of a silicon vertex detector (SVD),
a 50-layer central drift chamber (CDC),
an array of aerogel threshold Cherenkov counters (ACC), 
a barrel-like arrangement of
time-of-flight scintillation counters (TOF), 
and an electromagnetic calorimeter
comprised of CsI(Tl) crystals (ECL), all located inside
a superconducting solenoid coil that provides a 1.5~T magnetic field.
An iron flux-return located outside the coil is instrumented to 
detect $K_{\rm{L}}^0$ mesons and to identify muons (KLM).
The detector is described in detail elsewhere~\cite{Belle}.

Leptons are identified using likelihood ratios
calculated from the response of 
various subsystems of the detector.
For electron identification, the likelihood ratio is defined as 
${\cal P}(e) = {\cal{L}}_e/({\cal{L}}_e+{\cal{L}}_x)$,
where  ${\cal{L}}_e$ and ${\cal{L}}_x$ are the likelihoods 
for electron and non-electron hypotheses, respectively,
determined using the ratio of the energy deposit in the ECL to 
the momentum measured in the SVD and CDC, the shower shape in the ECL, 
the matching between the position 
of charged track trajectory and the cluster position in
the ECL, the hit information from the ACC and
the $dE/dx$ information in the CDC~\cite{EID}.
For muon  identification, the likelihood ratio is defined as 
(${\cal P}(\mu) = {\cal{L}_\mu}/({\cal{L}}_\mu+{\cal{L}}_{\pi}+{\cal{L}}_{K})$),
where  ${\cal{L}}_\mu$, ${\cal{L}}_\pi$  and ${\cal{L}}_K$ are the likelihoods 
for muon, pion and kaon hypotheses, respectively,
based on the matching quality and penetration depth of 
associated hits in the KLM~\cite{MUID}.

In order to optimize the event selection and to estimate
the signal efficiency,
we use Monte Carlo (MC) samples.
The signal and the background (BG) events from generic $\tau^+\tau^-$ decays are
generated by KORALB/TAUOLA~\cite{KKMC}. 
In the signal MC, we generate $\tau^+\tau^-$ pairs, where 
one $\tau$ decays into three leptons and the other $\tau$
decays generically.
All leptons from $\tau^-\to\ell^-\ell^+\ell^-$ decays are assumed to have a 
phase space distribution in the $\tau$ lepton's rest frame~\cite{XX}.
Other backgrounds including
$B\bar{B}$ and $e^+e^-\to q\bar{q}$ ($q=u,d,s,c$) processes,
Bhabhas, $e^+e^-\rightarrow\mu^+\mu^-$, 
and two-photon processes are generated by 
EvtGen~\cite{evtgen},
BHLUMI~\cite{BHLUMI}, 
KKMC~\cite{KKMC}, and
AAFHB~\cite{AAFH}, respectively. 
All kinematic variables are calculated in the laboratory frame
unless otherwise specified.
In particular,
variables
calculated in the $e^+e^-$ center-of-mass (CM) system
are indicated by the superscript ``CM''.

\section{Event Selection}

We search for $\tau^+\tau^-$ events in which one $\tau$ 
decays into three leptons~(signal $\tau$), 
while the other $\tau$ decays 
into  one charged track, any number of additional 
photons, and neutrinos~(tag $\tau$)\footnotemark[2].  
Candidate $\tau$-pair events are required to have 
four tracks with zero net charge.
\footnotetext[2]{Unless otherwise stated, 
charge-conjugate decays are implied throughout this paper.}
The following $\tau^-$ decays into three leptons are searched for:
$e^-e^+e^-$,
$\mu^-\mu^+\mu^-$,
$e^-\mu^+\mu^-$,
$\mu^-e^+e^-$,
$\mu^-e^+\mu^-$, and
$e^-\mu^+e^-$. 
Since each decay mode has a different mix of backgrounds,
the event selection is optimized mode by mode.
We optimize the selection criteria to improve the prospect of observing
evidence of a genuine signal, rather than to minimize the expected 
upper limits, as detailed later.

The event selection starts by reconstructing 
four charged tracks and any number of photons within the fiducial volume
defined by $-0.866 < \cos\theta < 0.956$,
where $\theta$ is the polar angle relative to 
the direction opposite to that of the incident $e^+$ beam in 
the laboratory frame.
The transverse momentum ($p_t$) of each charged track
and energy of each photon ($E_{\gamma}$) 
are required to satisfy the requirements
$p_t> $ 0.1 GeV/$c$ and $E_{\gamma}>0.1$ GeV, respectively.
For each charged track, the distance of the closest approach
with respect to the interaction point is required to be 
within $\pm$0.5 cm in the transverse direction 
and within $\pm$3.0 cm in the longitudinal direction.

%
%
The particles in an event are then separated into two 
hemispheres referred to as the signal and 
tag sides using the plane perpendicular to the thrust axis, as calculated from 
the observed tracks and photon candidates~\cite{thrust}.
The tag side contains a charged track
while the signal side contains three charged tracks.
We require all charged tracks on the signal side to be identified as leptons.
The electron (muon) identification criteria are ${\cal P}(e) > 0.9$ 
(${\cal P}(\mu) > 0.9$) for momenta
greater than 0.3 GeV/$c$ (0.6 GeV/$c$).
The electron (muon) identification efficiency
for our selection criteria is 91\% (85\%) 
while the probability of misidentifying a pion as an electron 
(muon) is below 0.5\% (2\%).

%
%
To ensure that the missing particles are neutrinos rather
than photons or charged particles that pass  outside the detector acceptance,
we impose additional requirements on the missing 
momentum $\vec{p}_{\rm miss}$,
which is calculated by subtracting the
vector sum of the momenta of all tracks and photons
from the sum of the $e^+$ and $e^-$ beam momenta.
We require that the magnitude of $\vec{p}_{\rm miss}$
be  greater than 0.4 GeV/$c$
and that its direction point into the fiducial volume of the
detector.

\begin{figure}
\begin{center}
\resizebox{0.5\textwidth}{0.5\textwidth}{\includegraphics{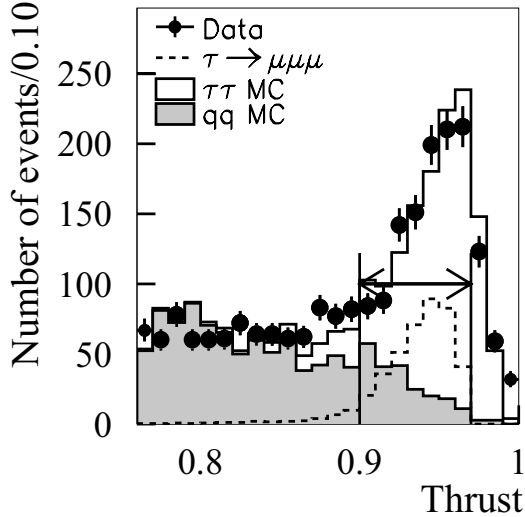}}
\caption{
Distribution of the thrust magnitude $T$ for the $\tau^- \to \mu^- \mu^+ \mu^-$
selection. The points with error bars are data, and 
the open histogram shows the BG estimated by MC.
The shaded histogram is the BG from $e^+e^- \to q \bar{q}$.
The dashed histogram is signal MC.
The region indicated by the arrow 
between the two vertical lines is selected.
}
\label{fig:thrust}
\end{center}
\end{figure}

%
%
To reject $q\bar{q}$ background,
the magnitude of thrust ($T$) should lie in the range
0.90 $< T <$ 0.97
for all modes except for $\tau^-\to e^-e^+e^-$
for which we require 0.90 $ < T  <$ 0.96.
The $T$ distribution for $\tau^- \to \mu^- \mu^+ \mu^-$ is shown in 
Fig.~\ref{fig:thrust}.
We also require $5.29$ GeV $< E^{\mbox{\rm{\tiny{CM}}}}_{\rm{vis}} < 9.5$ GeV, 
where $E^{\mbox{\rm{\tiny{CM}}}}_{\rm{vis}}$ 
is the total visible energy in the CM system, defined as 
the sum of the energies of the three leptons,
the charged track on the tag side (with a pion mass hypothesis)
and all photon candidates.

%
%
Since neutrinos are emitted only on the tag side,
the direction of $\vec{p}_{\rm miss}$
should lie within the tag side of the event.
The cosine of the opening angle between
$\vec{p}_{\rm miss}$ and the charged track on the tag side 
in the CM system, $\cos \theta^{\mbox{\rm \tiny CM}}_{\rm tag-miss}$, 
is therefore required to lie in the range 
$0.0<\cos \theta^{\mbox{\rm \tiny CM}}_{\rm tag-miss}<0.98$.
The requirement of 
$\cos \theta^{\mbox{\rm \tiny CM}}_{\rm tag-miss}<0.98$
suppresses Bhabha, $\mu^+\mu^-$ and two-photon backgrounds
since an undetected radiated photon 
results in a missing momentum in the same ECL cluster 
as the tag-side track~\cite{cite:tau_egamma}.
The reconstructed mass on
the tag side using a charged track (with a pion mass hypothesis) and photons,
$m_{\rm tag}$, is required to be less than 1.78 GeV/$c^2$.

%
%
Conversions ($\gamma\to e^+e^-$) are a large background for the 
$\tau^-\-\to e^-e^+e^-$ and $\mu^-e^+e^-$ modes.
We require $M_{ee}>0.2$ GeV/$c^2$, to reduce these
backgrounds further.
%
%
For the $\tau^-\to e^-e^+e^-$ and $\tau^-\to e^-\mu^+\mu^-$ modes,
the charged track on the tag side is required not to be an electron.
We apply the requirement
${\cal P}(e)<0.1$ since a large background 
from two-photon and Bhabha events still remains.
Furthermore, we reject the event if the projection of the
charged track on the tag side is in gaps between the ECL barrel and endcap.
%
%
To reduce backgrounds from Bhabha and $\mu^+\mu^-$ events
with extra tracks due to interaction 
with the detector material,
we require that the momentum in the CM system of 
the charged track on the tag side be less than 4.5 GeV/$c$
for the $\tau^-\to e^-e^+e^-$ and $\tau^-\to\mu^-e^+e^-$ modes.

%
%
Finally, 
to suppress  backgrounds from generic $\tau^+\tau^-$ and $q\bar{q}$ events, 
we apply a selection based on the magnitude of the missing momentum 
${p}_{\rm{miss}}$ 
and missing mass squared $m^2_{\rm{miss}}$ for all modes 
except for $\tau^-\to e^+\mu^-\mu^-$ and $\mu^+ e^-e^-$.
We do not apply this requirement for these two modes 
since their backgrounds are much smaller.
We apply different selection criteria depending on 
whether the $\tau$ decay on the tag side is  hadronic or leptonic:
the number of emitted neutrinos is two (one)
when the $\tau$ decay on the tag side is leptonic (hadronic).
Therefore, we separate events into two classes
according to the track on the tag side: leptonic or hadronic.
The selection criteria are listed in Table~\ref{tbl:misscut}.
\begin{table}
\begin{center}
\caption{
Selection criteria for 
the missing momentum ($p_{\rm{miss}}$) and
missing mass squared ($m^2_{\rm miss}$) for each mode.
The units for $p_{{\rm miss}}$ 
and  $m^2_{\rm miss}$ 
are GeV/$c$ and $({\rm{GeV}}/c^2)^2$, respectively.}
\label{tbl:misscut}
\begin{tabular}{c|l|l} \hline\hline
Mode & \multicolumn{1}{c|}{Hadronic tag} & \multicolumn{1}{c}{Leptonic tag} \\
\hline
$\tau^-\to\mu^-\mu^+\mu^-$ & $p_{\rm miss} > -3.0~m^2_{\rm miss}-1.0$ &
 $p_{\rm miss} > -2.5~m^2_{\rm miss}$ \\
$\tau^-\to \mu^- e^+ e^-$ &  $p_{\rm miss} > 3.0~m^2_{\rm miss}-1.5$  &
 $p_{\rm miss} > 1.3~m^2_{\rm miss}-1.0$ \\
$\tau^-\to e^- \mu^+ \mu^-$ & & \\ \hline
$\tau^-\to e^-e^+ e^-$ & $p_{\rm miss} > -3.0~m^2_{\rm miss}-1.0$ &
 $p_{\rm miss} > -2.5~m^2_{\rm miss}$ \\
 &  $p_{\rm miss} > 4.2~m^2_{\rm miss}-1.5$  &
 $p_{\rm miss} > 2.0~m^2_{\rm miss}-1.0$ \\ \hline
$\tau^-\to e^+ \mu^- \mu^-$ & \multicolumn{1}{c|}{not applied} & \multicolumn{1}{c}{not applied} \\
$\tau^-\to \mu^+ e^- e^-$ &    & \\ \hline\hline
\end{tabular}
\end{center}
\end{table}

%
%
For the optimization, we examine the relation
between the number of events 
($N_{\rm obs.}^{99}$), which would need to be observed
to obtain 99\% confidence level (CL) evidence,
and the number of expected BG events ($N_{\rm BG}$).
We find that better sensitivity is obtained for smaller $N_{\rm BG}$,
provided that the signal efficiency does not drop drastically.
For example, when we reduce $N_{\rm BG}$ from 1 to 0.1,
$N_{\rm obs.}^{99}$ decreases from 5 to 2, as 
calculated with the POLE program~\cite{pole}.
This is equivalent to an improvement of the effective 
efficiency by a factor of 2.5.

For the case of $\tau \to \mu \mu \mu$
we obtain an expected BG of $0.13\pm0.06$ with an efficiency of
7.6\% for the event selection described above. 
In this case, the branching fraction obtained from $N_{\rm obs.}^{99}$
is ${\cal B}_{99}= 1.8 \times 10^{-8}$, and the upper limit for the 
branching fraction at the 90\% CL is 
${\cal B}_{90}^{\rm UL}< 2.1 \times 10^{-8}$ for zero observed events.
When we relax the selection criteria by removing 
the requirements on
$p_{\rm miss}$-$m^2_{\rm miss}$, the momentum and mass of the tag side,
$\cos \theta_{\rm tag - miss}^{\rm CM}$, thrust, and so on,
we obtain an expected BG of $0.42\pm0.17$ with an efficiency of
8.9\%, so that ${\cal B}_{99}= 2.3 \times 10^{-8}$
and ${\cal B}_{90}^{\rm UL}< 1.6 \times 10^{-8}$ for zero 
observed events.
For the relaxed selection criteria, 
in the Feldman-Cousins approach~\cite{cite:FC}
the upper limits on the branching fractions are small
when the number of observed events fluctuates below
the number of expected background events.
As mentioned above, 
we optimize the selection criteria to obtain good sensitivity
for signal discovery. Therefore, we choose the selection criteria
described above to minimize ${\cal B}_{99}$
with the signal region defined below.

%
%
The following main background sources 
remain after the event selection, which
are estimated from the MC and data:
Bhabha and $\gamma\gamma \to e^+e^-$ for $\tau^- \to e^-e^+e^-$,
$\gamma\gamma \to \mu^+\mu^-$ for $\tau^- \to \mu^-e^+e^-$ and $e^-\mu^+\mu^-$,
$\tau$-pairs and $q \bar{q}$ for $\tau^- \to \mu^-\mu^+\mu^-$,
$e^-\mu^+e^-$ and $\mu^-e^+\mu^-$.

\section{Signal and Background Estimation}

The signal candidates are examined in two-dimensional plots 
of the $\ell^-\ell^+\ell^-$ invariant mass~($M_{\rm {3\ell}}$), and 
the difference between the summed energy and the 
beam energy in the CM system~($\Delta E$).
A signal event should have $M_{\rm {3\ell}}$
close to the $\tau$-lepton mass and $\Delta E$ close to zero.
We define an elliptical signal region in the $M_{\rm {3\ell}}$-$\Delta E$ plane,
which is optimized using the signal MC, 
to have a minimum area containing 90\% of the signal after
all the selections.

\begin{figure}
\begin{center}
\resizebox{0.4\textwidth}{0.4\textwidth}{\includegraphics{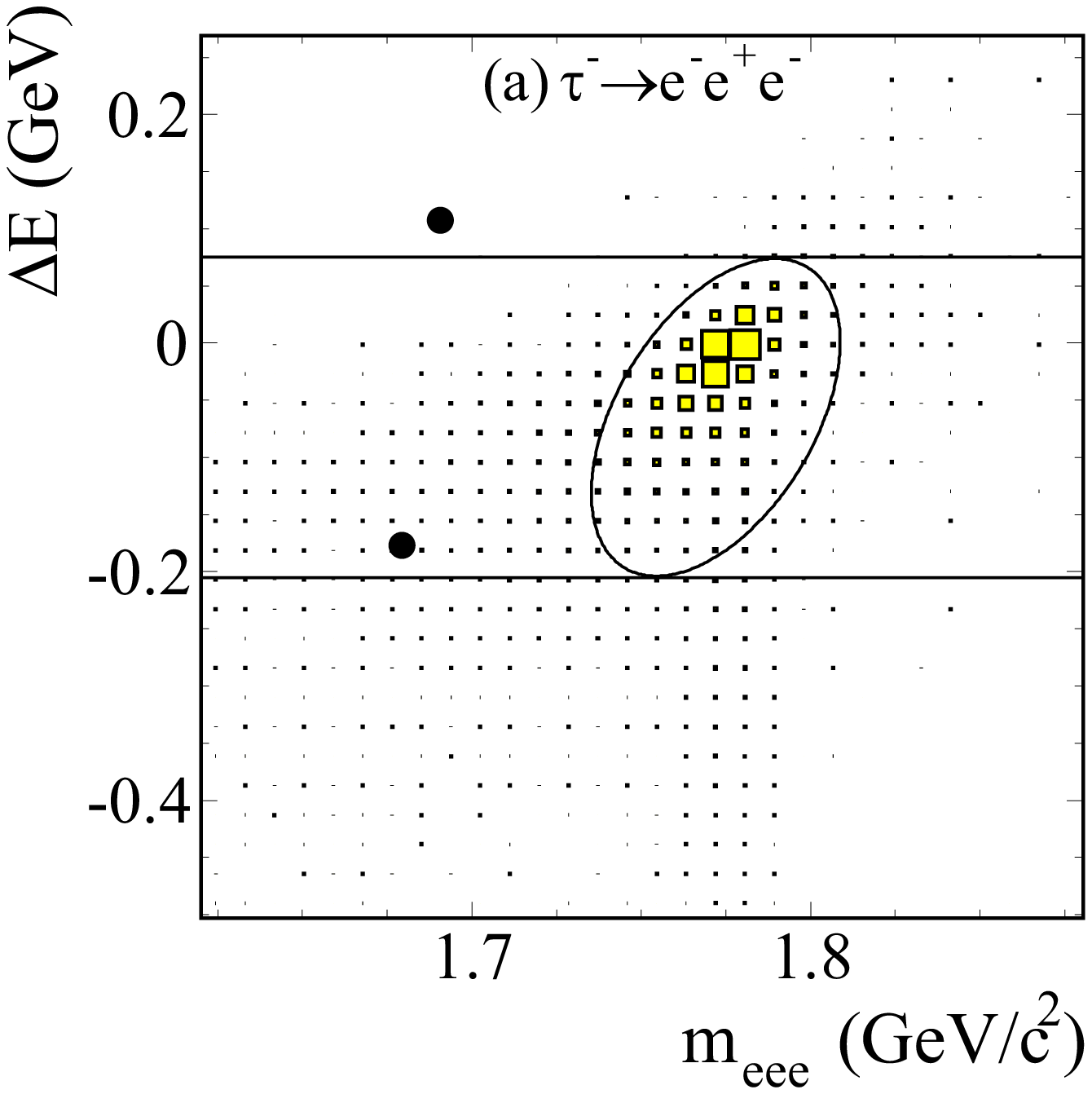}}
\resizebox{0.4\textwidth}{0.4\textwidth}{\includegraphics{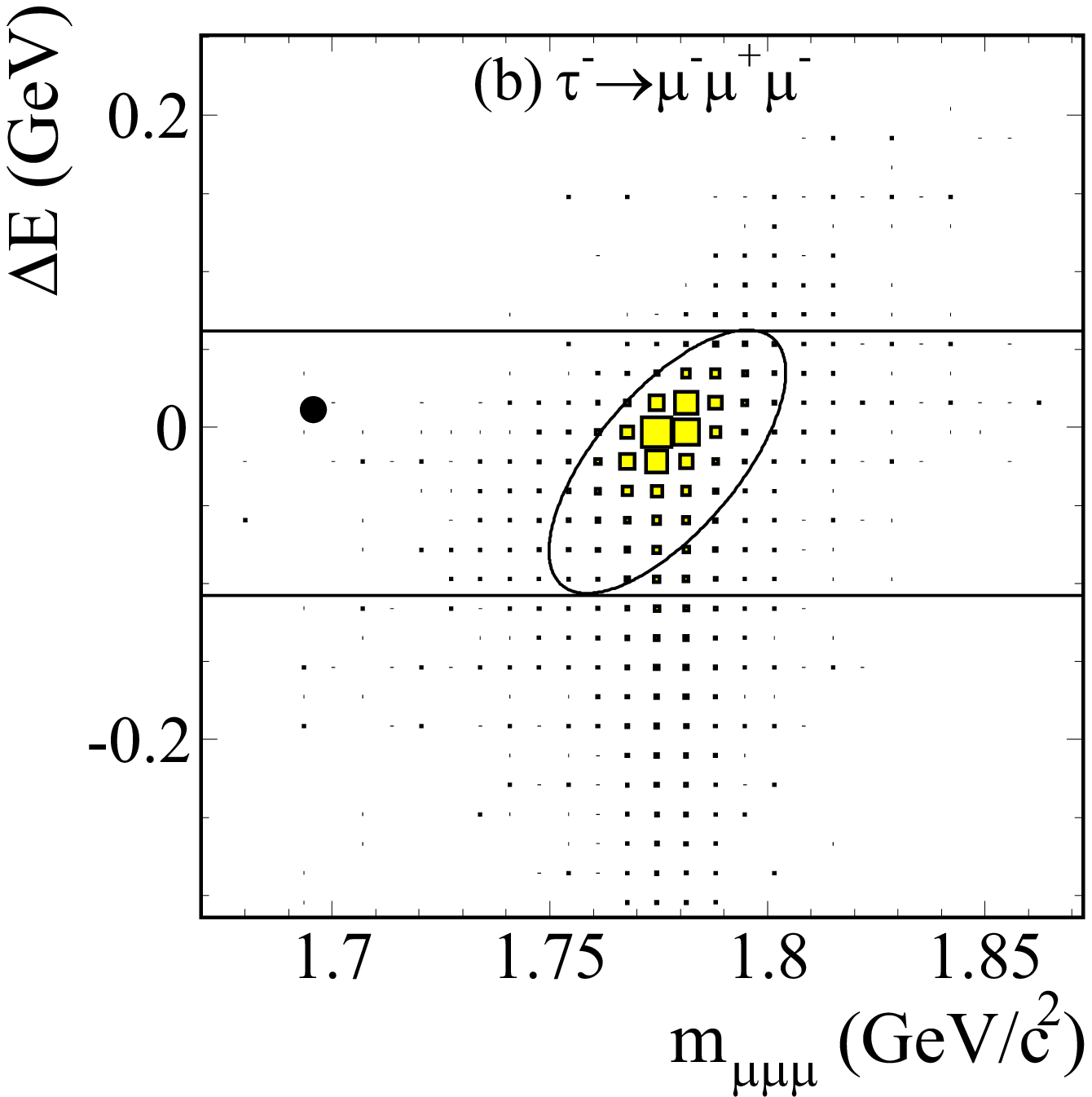}}\\
\resizebox{0.4\textwidth}{0.4\textwidth}{\includegraphics{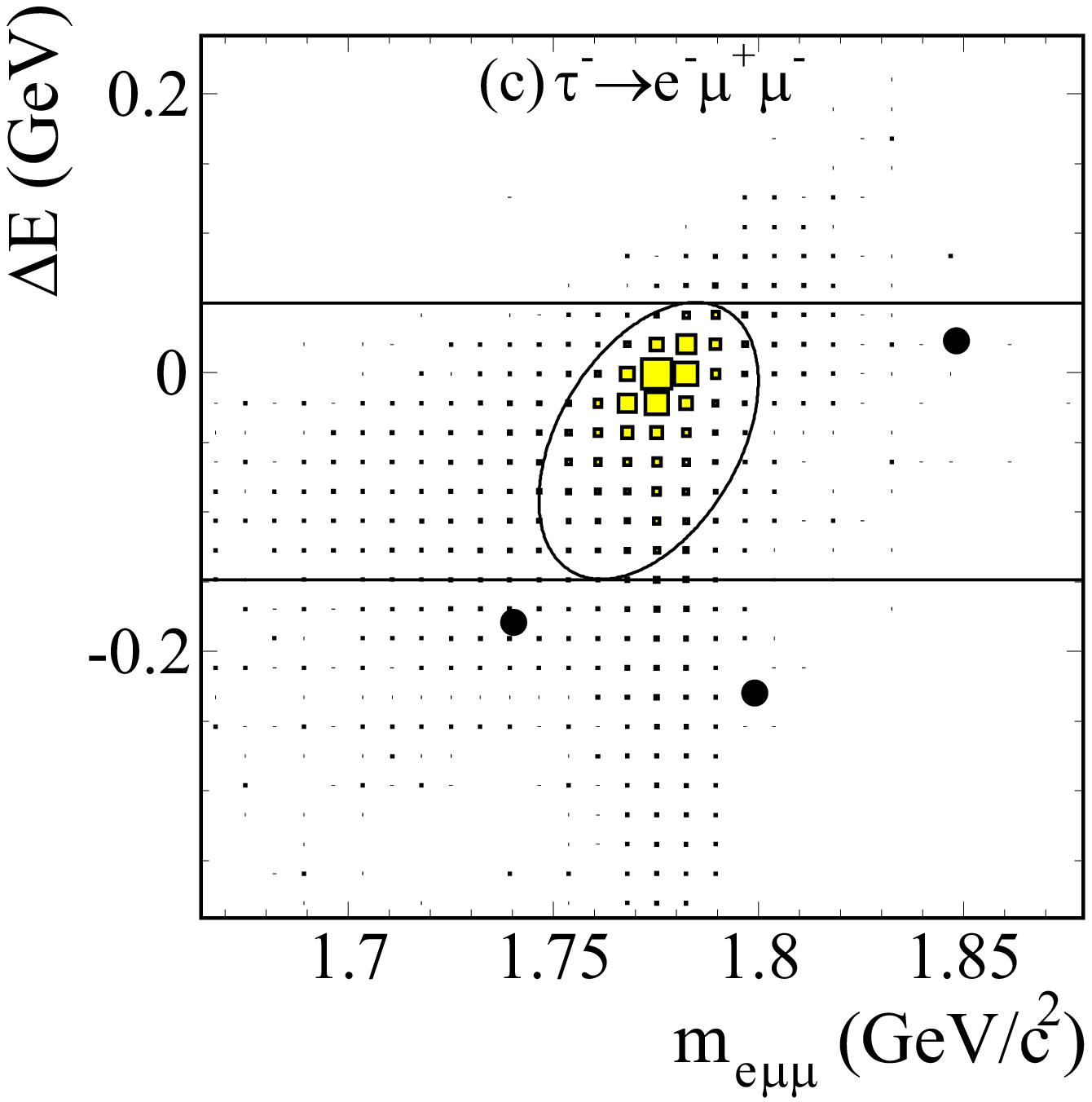}}
\resizebox{0.4\textwidth}{0.4\textwidth}{\includegraphics{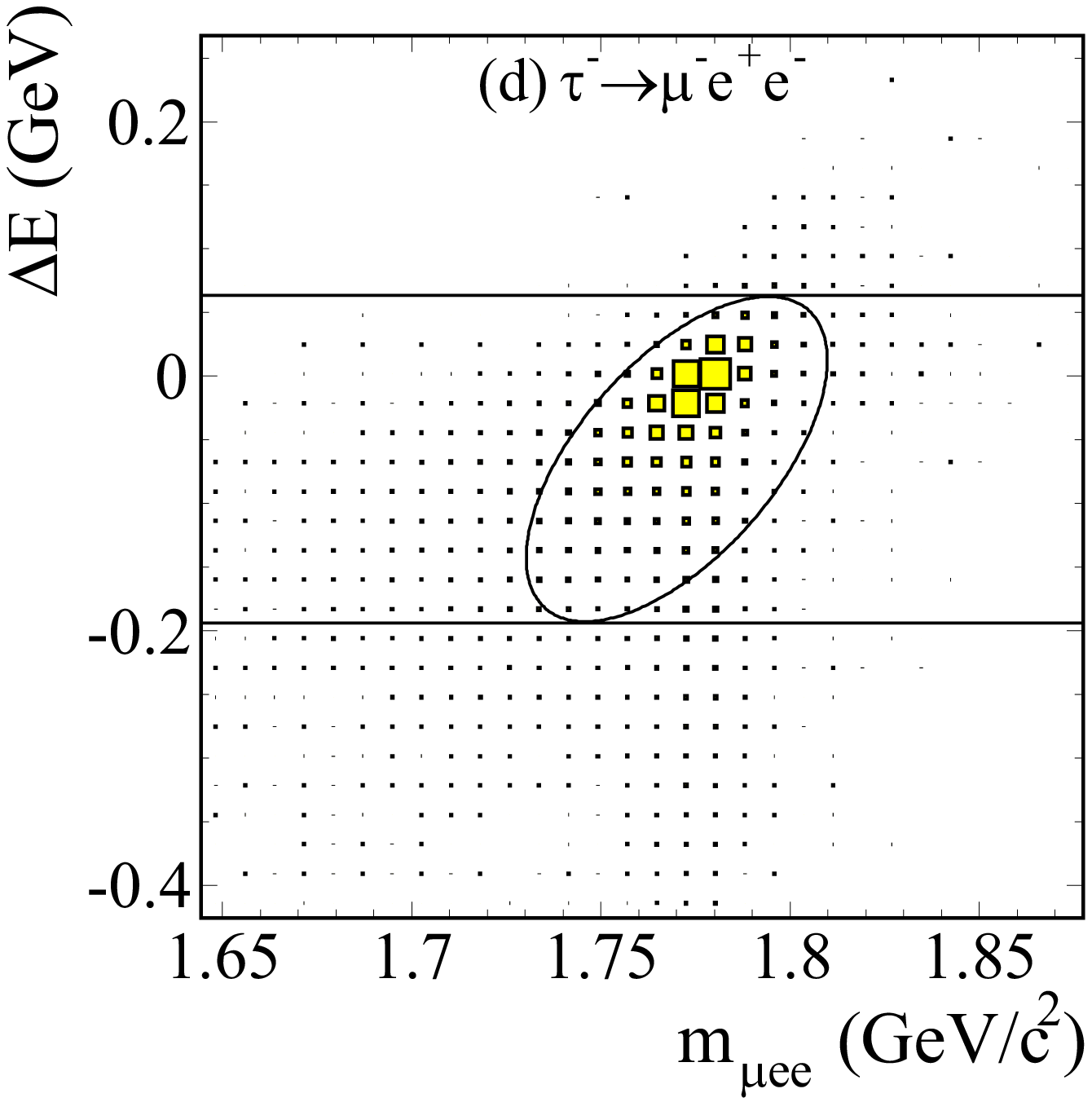}}\\
\resizebox{0.4\textwidth}{0.4\textwidth}{\includegraphics{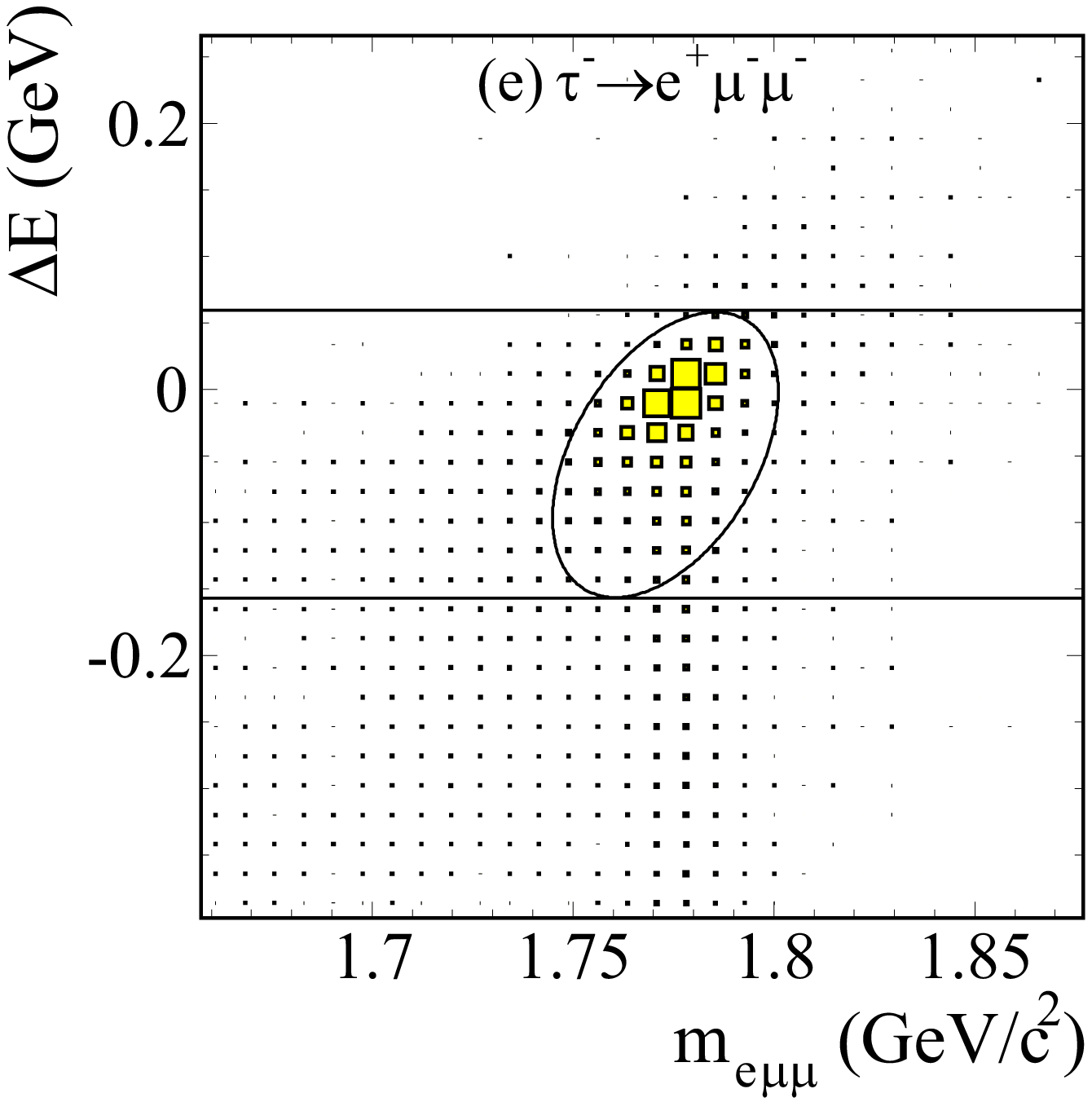}}
\resizebox{0.4\textwidth}{0.4\textwidth}{\includegraphics{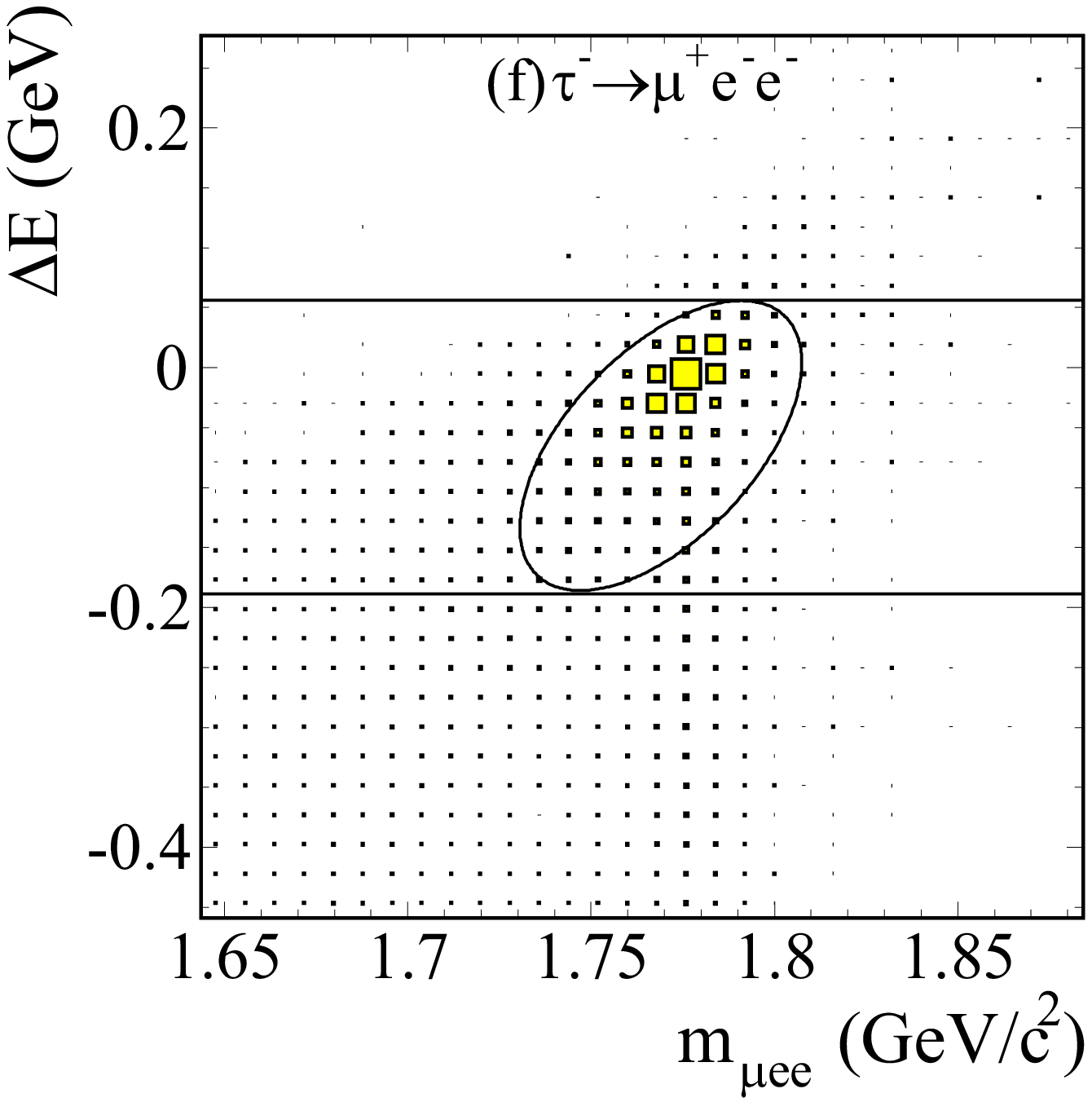}}
\caption{
Scatter-plots in the
$M_{3\ell}$ -- $\Delta{E}$ plane,
showing the $\pm 20 \sigma$ area for
(a) $\tau^-\rightarrow e^-e^+e^-$,
(b) $\tau^-\rightarrow\mu^-\mu^+\mu^-$,
(c) $\tau^-\rightarrow e^-\mu^+\mu^-$,
(d) $\tau^-\rightarrow\mu^- e^+e^-$,
(e) $\tau^-\rightarrow e^+\mu^-\mu^-$ and
(f) $\tau^-\rightarrow \mu^+e^-e^-$.
The data are indicated by the solid circles.
The filled boxes show the MC signal distribution
with arbitrary normalization.
The elliptical signal regions shown by the solid curves 
are used for evaluating the signal yield.
The region between the horizontal solid lines excluding
the signal region is
used to estimate the background expected in the elliptical region. 
}
\label{fig:openbox}
\end{center}
\end{figure}

In order not to bias our choice of selection criteria,
we blind the data in the signal region and estimate 
the signal efficiency and the number of background events from the MC 
and the data outside the signal region.
Figure~\ref{fig:openbox} shows scatter-plots
for the data and the signal MC distributed over $\pm 20\sigma$
on the $M_{\rm{3\ell}}-\Delta E$ plane.
We observe no events for $\tau^- \to \mu^-e^+e^-, e^-\mu^+e^-, \mu^-e^+\mu^-$,
one event for $\tau^- \to \mu^-\mu^+\mu^-$,
two events for $\tau^- \to e^-e^+e^-$ and
three events for $\tau^- \to e^-\mu^+\mu^-$,
outside the signal region.
The $\gamma$ conversion veto effectively reduces 
the background for $\tau^-\to e^-e^+e^-$.

The final estimate of the number of background events 
is based on the data with looser selection criteria
for particle identification and the event selection in the $M_{\rm{3\ell}}$ 
sideband region, which is defined as the box inside the horizontal lines but
excluding the signal region,
as shown by the horizontal lines in Fig.~\ref{fig:openbox}.
For example, we obtain 5 events in the sideband region for 
$\tau^- \to \mu^- \mu^+ \mu^-$, when a less stringent PID
criterion, ${\cal P}(\mu) > 0.6$, is applied.
Assuming that the background distribution is uniform in the sideband region, 
the number of background events in the signal box is estimated by 
interpolating the number of observed events in the sideband region 
into the signal region.
The signal efficiency and the number of expected background 
events for each mode are summarized in Table~\ref{tbl:eff}.

We estimate the systematic uncertainties due to lepton identification,
charged track finding, MC statistics, and  the integrated luminosity. 
The uncertainty due to the trigger efficiency is negligible compared 
with the other uncertainties.
The uncertainties due to lepton identification are
2.2\% per electron and 2.0\% per muon.
The uncertainty due to charged track finding is estimated to be 
1.0\% per charged track.
The uncertainty due to the electron veto on the tag side applied for the 
$\tau^-\to e^-e^+e^-$ and $\tau^-\to e^-\mu^+\mu^-$ modes is estimated 
to be the same as the uncertainty due to the electron identification.
For other modes, we use the same systematic uncertainty
for leptonic and hadronic decays on the tag side, 
because we do not apply any lepton/hadron identification 
requirements for any charged track on the tag side.
The uncertainties due to MC statistics and luminosity
are estimated to be (0.5 - 0.9)\% and 1.4\%, respectively.
We do not include an uncertainty due to the signal MC model.
All these uncertainties are added in quadrature, 
and the total systematic uncertainty for each mode is listed in 
Table~\ref{tbl:eff}.
\begin{table}
\begin{center}
\caption{
Results with nominal selection criteria: the
signal efficiency ($\varepsilon$), 
the number of expected background {events}  ($N_{\rm BG}$)
estimated from the sideband data, the total 
systematic uncertainty  ($\sigma_{\rm syst}$),
the number of observed events in the signal region ($N_{\rm obs}$)
and 90\% C.L. upper limit on the branching 
fraction~($\cal{B}$) for each individual mode. }
\label{tbl:eff}
\begin{tabular}{c|cccccc}\hline \hline
Mode &  $\varepsilon$~{(\%)} & 
$N_{\rm BG}$  & $\sigma_{\rm syst}$~{(\%)}
& $N_{\rm obs}$ & ${\cal{B}}(\times10^{-8})$ \\ \hline
$\tau^-\to e^-e^+e^-$      & 6.0 & 0.21$\pm$0.15 & 9.8 & 0 & $<$2.7 \\ 
$\tau^-\to\mu^-\mu^+\mu^-$ & 7.6 & 0.13$\pm$0.06 & 7.4 & 0 & $<$2.1 \\
$\tau^-\to e^-\mu^+\mu^-$  & 6.1 & 0.10$\pm$0.04 & 9.5 & 0 & $<$2.7 \\
$\tau^-\to \mu^-e^+e^-$    & 9.3 & 0.04$\pm$0.04 & 7.8 & 0 & $<$1.8 \\ 
$\tau^-\to e^+\mu^-\mu^-$  & 10.1& 0.02$\pm$0.02 & 7.6 & 0 & $<$1.7 \\
$\tau^-\to \mu^+e^-e^-$    & 11.5& 0.01$\pm$0.01 & 7.7 & 0 & $<$1.5 \\ 
\hline\hline
\end{tabular}
\end{center}
\end{table}

\section{Upper Limits on the branching fractions}

Finally, we examine the signal region and find no events
for all considered modes. Therefore,
we set upper limits on the branching fractions 
of $\tau^-\to\ell^-\ell^+\ell^-$
based on the Feldman-Cousins method.
The 90\% C.L. upper limit on the number of signal events 
including the systematic uncertainty~($s_{90}$) 
is obtained by the POLE program without conditioning~\cite{pole} with 
the number of expected background events, the number of observed events
and the systematic uncertainty.
The upper limit on the branching fraction ($\cal{B}$) is then given by
\begin{equation}
{{\cal{B}}(\tau^-\to\ell^-\ell^+\ell^-) <
\displaystyle{\frac{s_{90}}{2N_{\tau\tau}\varepsilon{}}},}
\end{equation}
where the number of $\tau$ pairs, $N_{\tau\tau} =  719\times 10^6$,
is obtained from the integrated luminosity of 782 fb$^{-1}$ and
the cross section of $\tau$ pair production, which 
is calculated in the updated version of KKMC~\cite{tautaucs} to be 
$\sigma_{\tau\tau} = (0.919 \pm 0.003)$ nb.
The 90\% C.L. upper limits on the branching fractions 
${\cal{B}}(\tau^-\rightarrow \ell^- \ell^+\ell^-)$  are in the range
between $1.5 \times 10^{-8}$ and $2.7 \times 10^{-8}$
and are summarized in Table~\ref{tbl:eff}.

\section{Summary}

We report results of a search for lepton-flavor-violating $\tau$ decays 
into three leptons using 782 fb$^{-1}$ of data.
No events are observed and we set 90\% C.L. upper limits 
on the branching fractions:
${\cal{B}}(\tau^-\rightarrow e^-e^+e^-) < 2.7\times 10^{-8}$, 
${\cal{B}}(\tau^-\rightarrow \mu^-\mu^+\mu^-) < 2.1\times 10^{-8}$, 
${\cal{B}}(\tau^-\rightarrow e^-\mu^+\mu^-) < 2.7\times 10^{-8}$, 
${\cal{B}}(\tau^-\rightarrow \mu^-e^+e^-) < 1.8\times 10^{-8}$, 
${\cal{B}}(\tau^-\rightarrow e^+\mu^-\mu^-) < 1.7\times 10^{-8}$
and  
${\cal{B}}(\tau^-\rightarrow \mu^+e^-e^-) < 1.5\times 10^{-8}$.
These results improve the best previously published upper limits
by factors from 1.3 to 1.6, and are the most
stringent upper limits of all $\tau$ decays.
These upper limits can be used
to constrain the space of parameters in various models beyond the SM.

\section*{Acknowledgments}

%
We are grateful to M.J.~Herrero for useful discussions.
We thank the KEKB group for the excellent operation of the
accelerator, the KEK cryogenics group for the efficient
operation of the solenoid, and the KEK computer group and
the National Institute of Informatics for valuable computing
and SINET3 network support.  We acknowledge support from
the Ministry of Education, Culture, Sports, Science, and
Technology (MEXT) of Japan, the Japan Society for the 
Promotion of Science (JSPS), and the Tau-Lepton Physics 
Research Center of Nagoya University; 
the Australian Research Council and the Australian 
Department of Industry, Innovation, Science and Research;
the National Natural Science Foundation of China under
contract No.~10575109, 10775142, 10875115 and 10825524; 
the Department of Science and Technology of India; 
the BK21 and WCU program of the Ministry Education Science and
Technology, the CHEP SRC program and Basic Research program (grant No.
R01-2008-000-10477-0) of the Korea Science and Engineering Foundation,
Korea Research Foundation (KRF-2008-313-C00177),
and the Korea Institute of Science and Technology Information;
the Polish Ministry of Science and Higher Education;
the Ministry of Education and Science of the Russian
Federation and the Russian Federal Agency for Atomic Energy;
the Slovenian Research Agency;  the Swiss
National Science Foundation; the National Science Council
and the Ministry of Education of Taiwan; and the U.S.\
Department of Energy.
This work is supported by a Grant-in-Aid from MEXT for 
Science Research in a Priority Area ("New Development of 
Flavor Physics"), and from JSPS for Creative Scientific 
Research ("Evolution of Tau-lepton Physics").


%
%
%

\end{document}